\documentclass[manuscript]{acmart}

\AtBeginDocument{%
  \providecommand\BibTeX{{%
    \normalfont B\kern-0.5em{\scshape i\kern-0.25em b}\kern-0.8em\TeX}}}

\usepackage{booktabs} 
\usepackage{float}
\usepackage{caption}
\usepackage{subcaption}

\usepackage{xspace}
\newcommand{\idest}{i.e.,\xspace}
\newcommand{\eg}{e.g.,\xspace}

\newcommand{\pbeta}{\ensuremath{\text{RP}^3\!\beta}\xspace}
\newcommand{\pbetabold}{\ensuremath{\text{\textbf{RP}}^\textbf{3}\!{\boldsymbol \beta}}\xspace}
\newcommand{\EASER}{EASE$^R$\xspace}

\setcopyright{rightsretained} 

\begin{document}

\copyrightyear{2021} 
\acmYear{2021} 
\acmConference[IMX '21]{ACM International Conference on Interactive Media Experiences}{June 21--23, 2021}{Virtual Event, NY, USA}
\acmBooktitle{ACM International Conference on Interactive Media Experiences (IMX '21), June 21--23, 2021, Virtual Event, NY, USA}
\acmDOI{10.1145/3452918.3465493}
\acmISBN{978-1-4503-8389-9/21/06}

\title{Measuring the User Satisfaction in a Recommendation Interface with Multiple Carousels}

\author{Nicol\`o Felicioni}
\orcid{0000-0002-3555-7760}
\affiliation{%
  \institution{Politecnico di Milano}
  \country{Italy}
}
\email{nicolo.felicioni@polimi.it}

\author{Maurizio Ferrari Dacrema}
\orcid{0000-0001-7103-2788}
\affiliation{%
  \institution{Politecnico di Milano}
  \country{Italy}
}
\email{maurizio.ferrari@polimi.it}

\author{Paolo Cremonesi}
\orcid{0000-0002-1253-8081}
\affiliation{%
  \institution{Politecnico di Milano}
  \country{Italy}
  }
\email{paolo.cremonesi@polimi.it}



\begin{abstract}
It is common for video-on-demand and music streaming services to adopt a user interface composed of several recommendation lists, \idest \emph{widgets} or \emph{swipeable carousels}, each generated according to a specific criterion or algorithm (\eg most recent, top popular, recommended for you, editors' choice, etc.). 
Selecting the appropriate combination of carousel has significant impact on user satisfaction.
A crucial aspect of this user interface is that to measure the relevance a new carousel for the user it is not sufficient to account solely for its individual quality. Instead, it should be considered that other carousels will already be present in the interface. 
This is not considered by traditional evaluation protocols for recommenders systems, in which each carousel is evaluated in isolation, regardless of (i) which other carousels are displayed to the user and (ii) the relative position of the carousel with respect to other carousels. 
Hence, we propose a two-dimensional \emph{evaluation protocol for a carousel setting} that will measure the quality of a recommendation carousel based on how much it improves upon the quality of an already available set of carousels.
Our evaluation protocol takes into account also the position bias, \idest users do not explore the carousels sequentially, but rather concentrate on the top-left corner of the screen.

We report experiments on the movie domain and notice that under a carousel setting the definition of which criteria has to be preferred to generate a list of recommended items changes with respect to what is commonly understood. 
\end{abstract}



\maketitle

\section{Introduction}
Recommender systems have become a widespread technology, which is present in a multitude of the services we use today. Their main goal is to help the user explore the vast catalogs at their disposal and are very important tools to ensure and improve the user satisfaction. 
The ability to provide high quality personalized recommendation is central in many business models, among which video on demand and music streaming services. Most of the research that proposes new recommendation algorithms is evaluated with offline experiments which are known to be often insufficient to accurately predict user satisfaction due to a number of factors \cite{10.1145/2532508.2532511}. 


In a traditional \textit{offline} evaluation scenario, the systems is asked to generate a list of recommended items to a set of users, one list for each user. 
Each list is than compared with the historical opinions of the user and quality is measured with standard accuracy metrics (such as NDCG) developed in the fields of information retrieval and machine learning. 

\begin{figure*}[]
    \centering
    \begin{minipage}{.49\textwidth}
        \centering
        \includegraphics[width=1\textwidth]{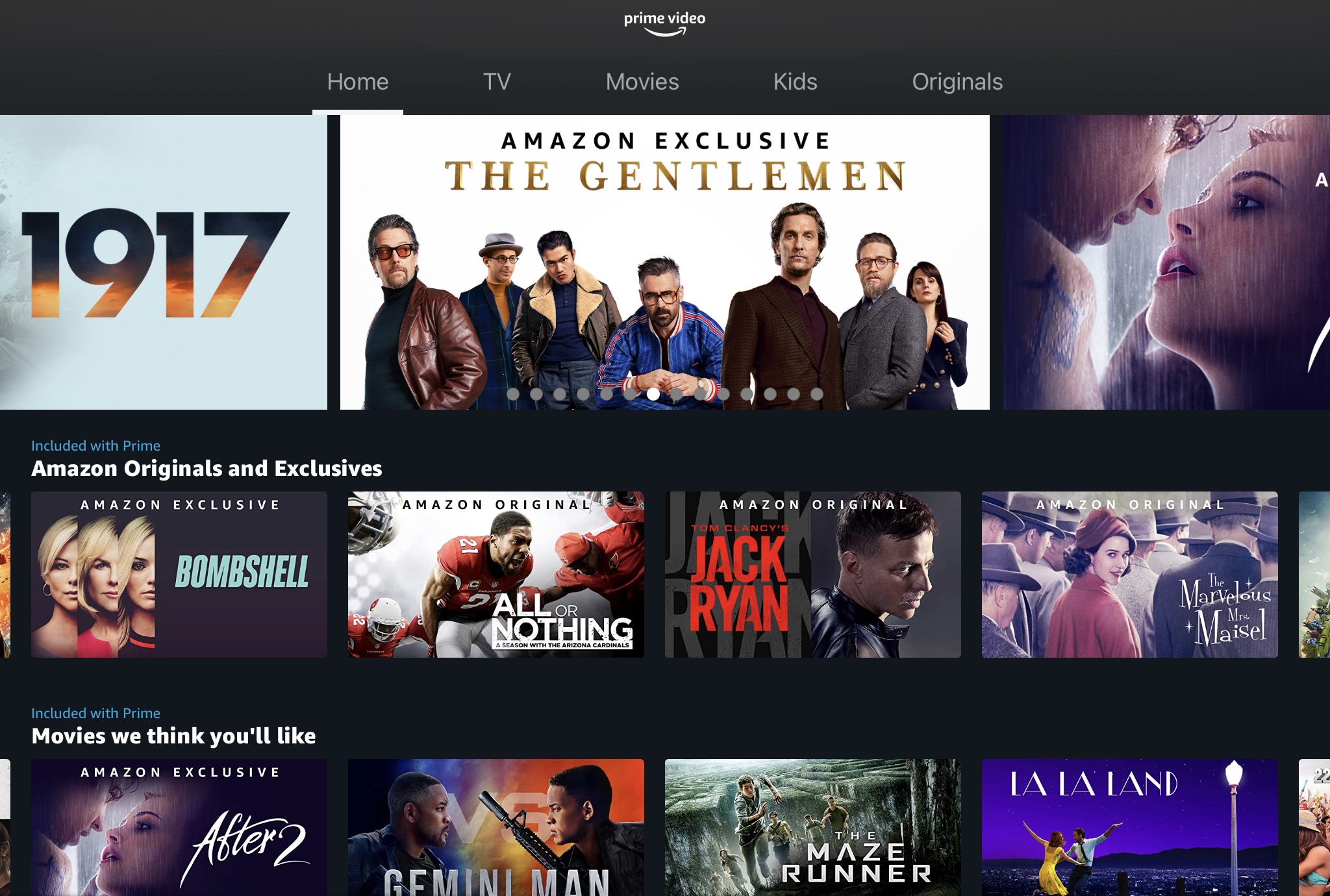}
    \end{minipage}%
    \hfill
    \begin{minipage}{0.49\textwidth}
        \centering
        \includegraphics[width=1.\textwidth]{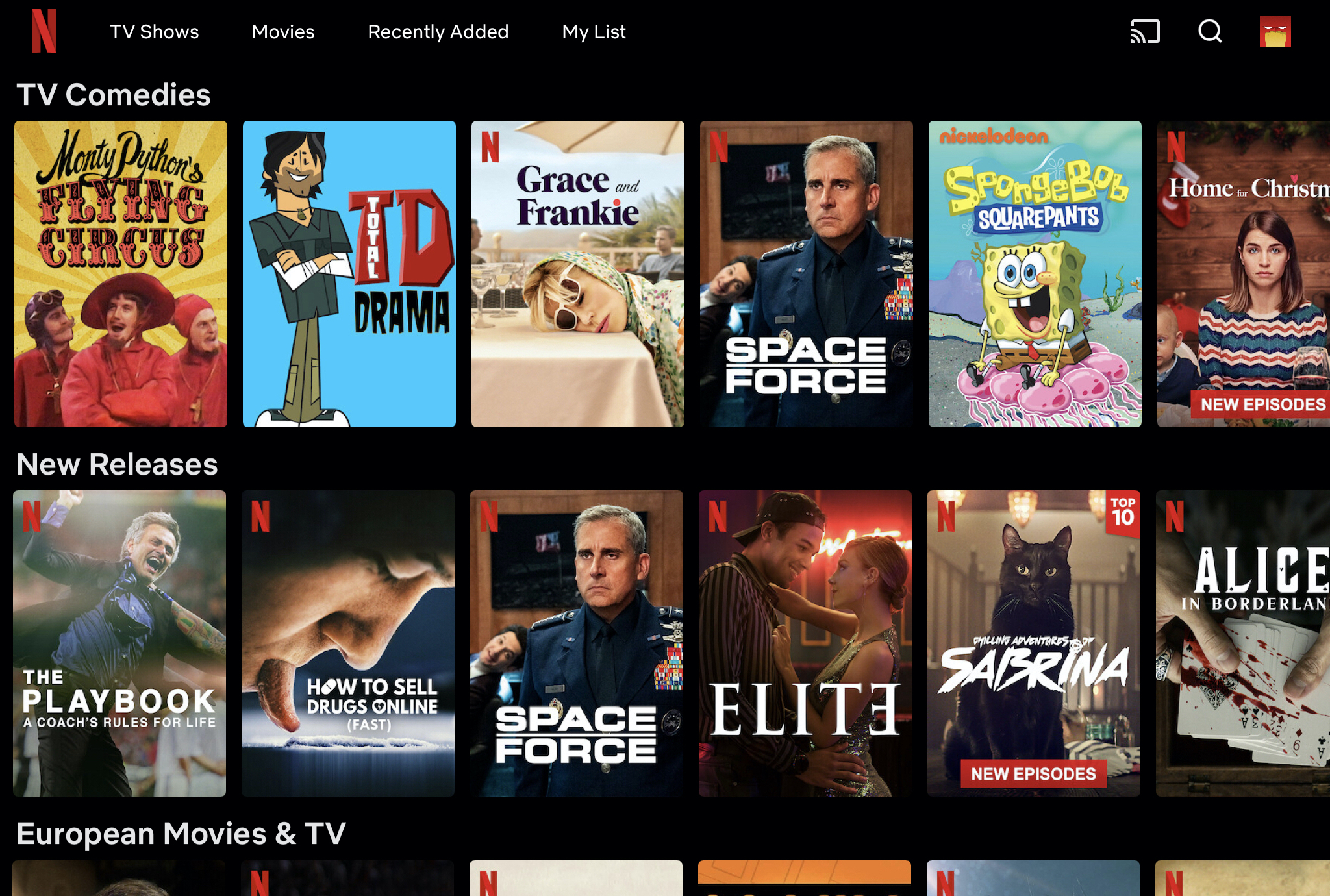}
    \end{minipage}
    \caption{Examples of carousel user interfaces in the multimedia streaming domain. On the left, there is the Amazon Prime Video homepage, while on the right, there is the Netflix homepage.}
    \Description{In this figure there are two examples of carousel user interfaces taken from real applications.}
    \label{fig:amazon_netflix}
\end{figure*}

\begin{figure*}[]
    \centering
    \begin{minipage}{0.45\textwidth}
        \centering
        \includegraphics[width=0.7\textwidth]{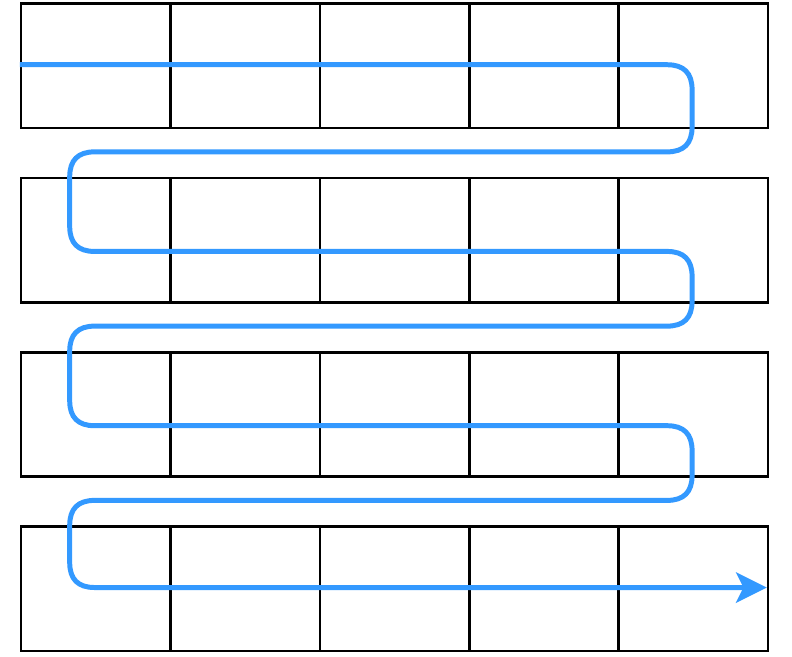}
        \caption{Traditional offline evaluation strategy in which all carousels are concatenated, assuming the user will explore them sequentially one at a time.}
        \Description{In this Figure, there is a drawing of a generic carousel interface with an arrow that represents the user exploration of the interface. In this case, the arrow simulates the assumption that the user will explore the carousels sequentially one at a time. The exploration will start from the first carousel on top and the first element on the left, and continue on the right. Then, the exploration of the second carousel will start again from the first element on the left and continue on the right. This is the assumption made by traditional offline evaluation metrics.} 
        \label{fig:concat-list-eval}
    \end{minipage}
        \hfill
    \begin{minipage}{.45\textwidth}
        \centering
        \includegraphics[width=0.7\textwidth]{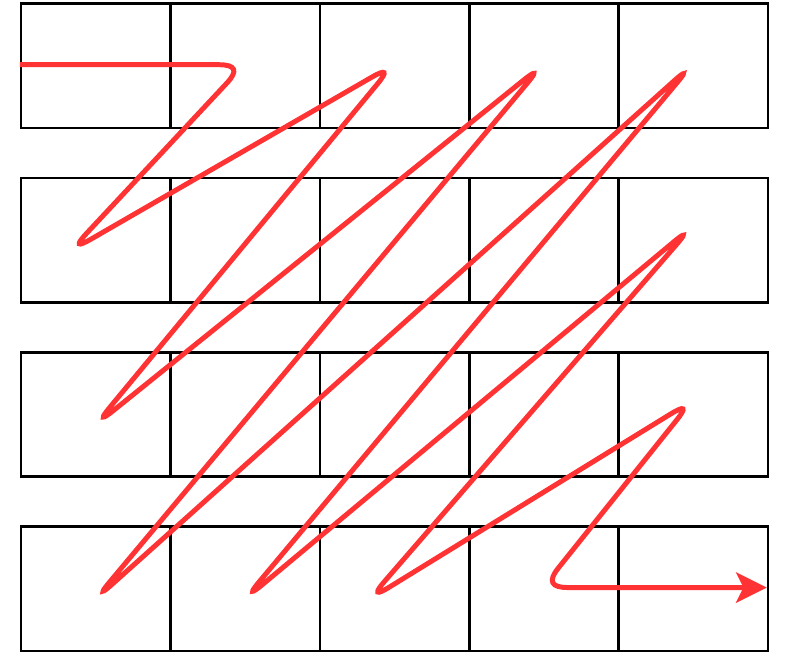}
        \caption{The proposed 2-D offline evaluation strategy which accounts for the way users explore a two-dimensional recommendation carousel by concentrating on the top-left corner and then proceeding both to the right and to the bottom across different carousels.}
        \Description{In this Figure, there is a drawing of a generic carousel interface with an arrow that represents the user exploration of the interface. In this case, the arrow simulates the assumption that the user will explore a two-dimensional interface focusing first on the top-left corner, with decreasing attention both to the right and to the bottom.
        This is the assumption made by the newly proposed metric.} 
        \label{fig:2d-list-eval}
    \end{minipage}%

\end{figure*}

This approach, however, disregards an essential component of the user experience, that is the user interface. In fact, in video-on-demand services (\eg Netflix, Amazon Prime Video) and music streaming platforms (\eg Spotify) users are frequently provided with several recommendation lists or \emph{carousels}, each with a certain theme \eg recently added, originals, trending, editorially curated (see Figure \ref{fig:amazon_netflix}).
In this scenario, the user satisfaction depends on all the carousels that are shown, not just one, and there is significant industrial interest in finding effective strategies to select which carousels to display \cite{DBLP:conf/recsys/BendadaSB20,DBLP:conf/kdd/DingGV19,DBLP:conf/recsys/WuASB16}. In order to represent such a scenario more closely, it is necessary to take into account how the recommendations in the various carousels complement each other and how the user explore a two-dimensional interface.
There seems to be no agreed evaluation protocol to address both cases. Sometimes the recommendation lists are simply concatenated one after the other, or the average of the recommendation quality is computed (see Figure \ref{fig:concat-list-eval}). Alternatively, the evaluation is done not in terms of correct recommendations, but rather in terms of which carousel contains at least a correct recommendation \cite{DBLP:conf/recsys/WuASB16}. These approaches we argue are insufficient because they do not take into account how the two-dimensional structure of the page affects the user behavior and how the recommendation lists complement each other. Consider, for example, a scenario where there is a page containing a carousel and we need to select the second one. Clearly, we should not select as second carousel one which contains recommendations very similar to the first one (even if it has, taken individually, a high recommendation quality), rather we should balance the recommendation quality and diversity of the recommendations \cite{DBLP:conf/recsys/ZhangH08}. In particular, we should compute the recommendation quality of the \emph{second} carousel by accounting only for the \emph{new} correct recommendations.
Furthermore, to the best of our knowledge no metric exists that measures the ranking quality of the recommendations based on the way users tend to explore a two-dimensional interface, by focusing on the top-left corner of the screen and then exploring the recommendations both to the right and to the bottom. 
We also note that most of the few articles discussing recommendation strategies for carousel-based user interfaces are evaluated in an online setting (e..g, with surveys or A/B testing) and the absence of a standardized offline evaluation methodology makes it difficult for researchers to explore a carousel-based user interface.

In this paper we propose a definition for this carousel setting highlighting the specific characteristics that distinguish both the user interface and the user behavior from the traditional single list evaluation. 
We also propose a 2-D strategy for the offline evaluation tailored for a carousel setting as well as a ranking metric which takes into account how users explore such interfaces (see Figure \ref{fig:2d-list-eval}). 
We finally report experimental results indicating that accounting for the carousel interface and user behavior changes the relative performance of recommendation models such that they lead to a different conclusion on which should be the carousel to display.



The rest of the paper is organized as follows. In Section \ref{sec:related_works} we presents related works on carousel interfaces, in Section \ref{sec:evaluation_protocol} we define and describe the carousel scenario as well as
the two-dimensional ranking metric.
Finally, in Section \ref{sec:experimental_anaysys} we report the results of our experimental analysis and in Section \ref{sec:conclusions} we draw the conclusions and possible future works.

\section{Related Work}
\label{sec:related_works}

\begin{figure*}[]
    \centering
    \begin{minipage}{0.49\textwidth}
        \centering
        \includegraphics[width=0.9\textwidth]{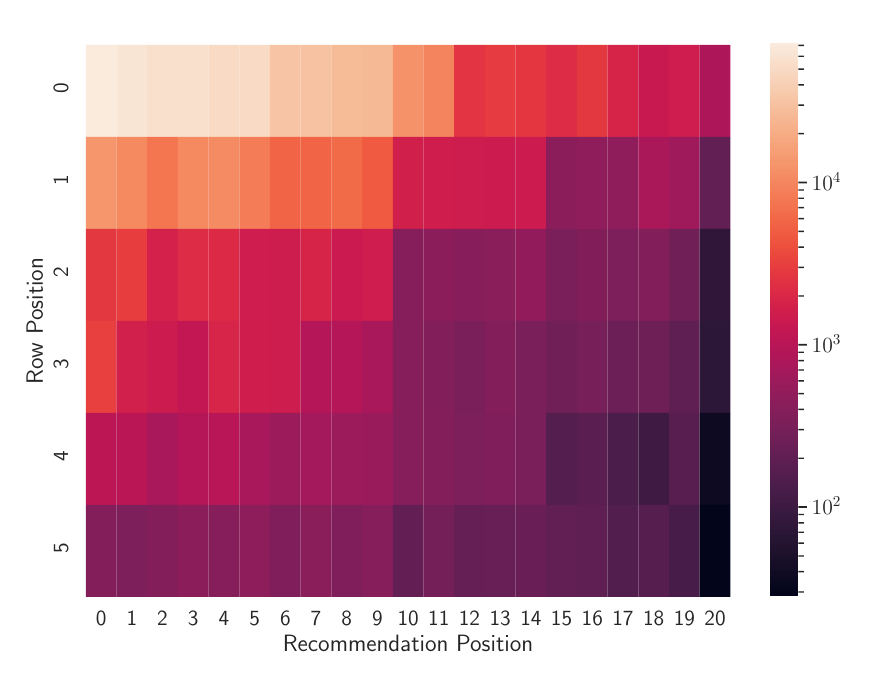}
        \caption{Heatmap of the number of interactions per position on the screen taken from the dataset presented in \cite{DBLP:conf/cikm/MaureraDSSC20}. Values are log-scaled.}
        \Description{In this figure, we can see a heatmap of the interactions made with a carousel interface in a real-world scenario. We can see how the interactions decrease both to the right and to the bottom, consistently with prior research in the field.}
        \label{fig:cw_heatmap}
    \end{minipage}
        \hfill
        \begin{minipage}{.49\textwidth}
        \centering
        \includegraphics[width=0.60\textwidth]{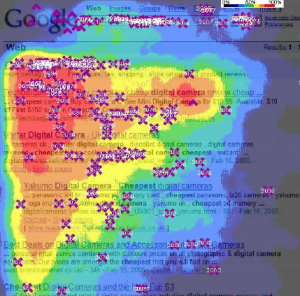}
        \caption{A known visualization of how users navigate a simple search result page, the attention is focused on a top-left \emph{golden triangle} and progressively decreases moving to the right and the bottom \cite{DBLP:conf/wsdm/ChierichettiKR11}.
        }
        \Description{In this figure, we can see a heatmap of the interactions made with a search result page on a web search engine. Also here, we can see how the interactions decrease both to the right and to the bottom. This effect is the so-called golden triangle.}
        \label{fig:ir_positionbias}
    \end{minipage}%

\end{figure*}

Many research works tried to understand user behavior in a two-dimensional setting, usually by employing eye-tracking.
For example, \citet{DBLP:conf/etra/KammererG10} examine the differences in user behavior when presented the same 9 results of a search engine result page in two different arrangements. The page is displayed as a plain list to some of the users involved, while it is displayed as a 3x3 two-dimensional grid to others. The study indicated that users navigate the list sequentially from top to bottom, while the grid navigation traverses both rows and columns according to a two-dimensional pattern. 
\citet{DBLP:conf/recsys/ZhaoCHK16} analyze the behavior of users of a two-dimensional recommendation engine. They make the examples of YouTube, MovieLens, Hulu (three video streaming services), and Google Apps (a popular mobile application store). 
From their analyses, the authors conclude that there is an "F-pattern" of the users' gaze behavior.
In other words, users are more focused on the top-left corner of a two-dimensional interface and that the attention is lower both to the bottom and to the right. This phenomenon is also called the "golden triangle" and it can be visualized by looking at Figure \ref{fig:ir_positionbias}.

While there are industrial research papers focusing on the carousel scenario, indicating the importance of this topic for the industry, there seems to be much less papers from the academia. This may be due to the requirement that such studies be evaluated in an online setting and to the lack of an offline evaluation protocol for the carousel setting. Such an evaluation protocol could allow researchers to more easily work on the topic even when they do not have easy access to a real user base.
\citet{DBLP:conf/recsys/WuASB16} try to optimize the ranking of the carousels on the user interface of Netflix, a popular online video streaming service. In this work, they focus on a carousel interface that Netflix uses, with a fixed recommendation on top of the screen and a set of thematically coherent carousels.
Their goal is not to design a novel UI layout, but to include navigation feedback to understand better the user satisfaction on their services.
\citet{DBLP:conf/wsdm/GrusonCCMHTC19} optimize the homepage of Spotify, an online music streaming service.
In this work, they focus on a carousel interface visualized on a mobile device, with users that can scroll left and right within carousels to reveal more items, and up and down to reveal more carousels.
In their offline evaluation, the carousels are sequentially concatenated as a single long recommendation list. 
Evaluating in this way, though, does not reflect the actual user behavior in a two-dimensional layout.
Also, the authors perform an online evaluation and find discrepancies between online and offline evaluations.
\citet{DBLP:conf/recsys/BendadaSB20} try to optimize a carousel layout of the homepage of Deezer, a music streaming service. In particular, they design a real-time model to understand the user preferences and adjust the carousels displayed. 
\citet{DBLP:conf/kdd/DingGV19} target the problem of optimizing the carousel homepage of Amazon Prime Video, a video streaming service. 
They point out that business constraints may fix carousels in certain positions of the grid, so the evaluation should take into account how the other carousel intersect with the fixed ones and how they improve the user experience.

\section{Evaluation in a Carousel Layout}
\label{sec:evaluation_protocol}

In this section we describe common characteristics of a real industrial service that employs a carousel layout.

First of all, the \textit{interface} is two-dimensional. The user is presented with a grid of recommendations, divided in carousels, which are the rows of this grid. Each carousel may be generated by different providers. In general the provider of a certain carousel is not aware of the recommendations contained in the others. This means that among different carousels there may be duplicate recommendations\footnote{For example, in the Netflix homepage shown in Figure \ref{fig:amazon_netflix} the TV series \emph{Space Force} appears both in the \emph{TV Comedies} and \emph{New Releases} carousels.}, while inside a single carousel this is avoided.

In terms of \textit{user behavior}, it is known that users do not explore a grid of recommendations sequentially, one carousel at a time, but instead they concentrate on the top-left corner. This should be taken into account when calculating the quality of the provided ranking.

Furthermore, in this scenario a content provider will need to account for as much information as possible on the user behavior and interface characteristics in order to be able to select which are the best recommendation carousels to display.

\subsection{NDCG in two dimensions}
\label{sec:evaluation_metrics}

\begin{figure*}
    \centering
    \begin{subfigure}{.48\textwidth}
    \includegraphics[width=0.9\columnwidth]{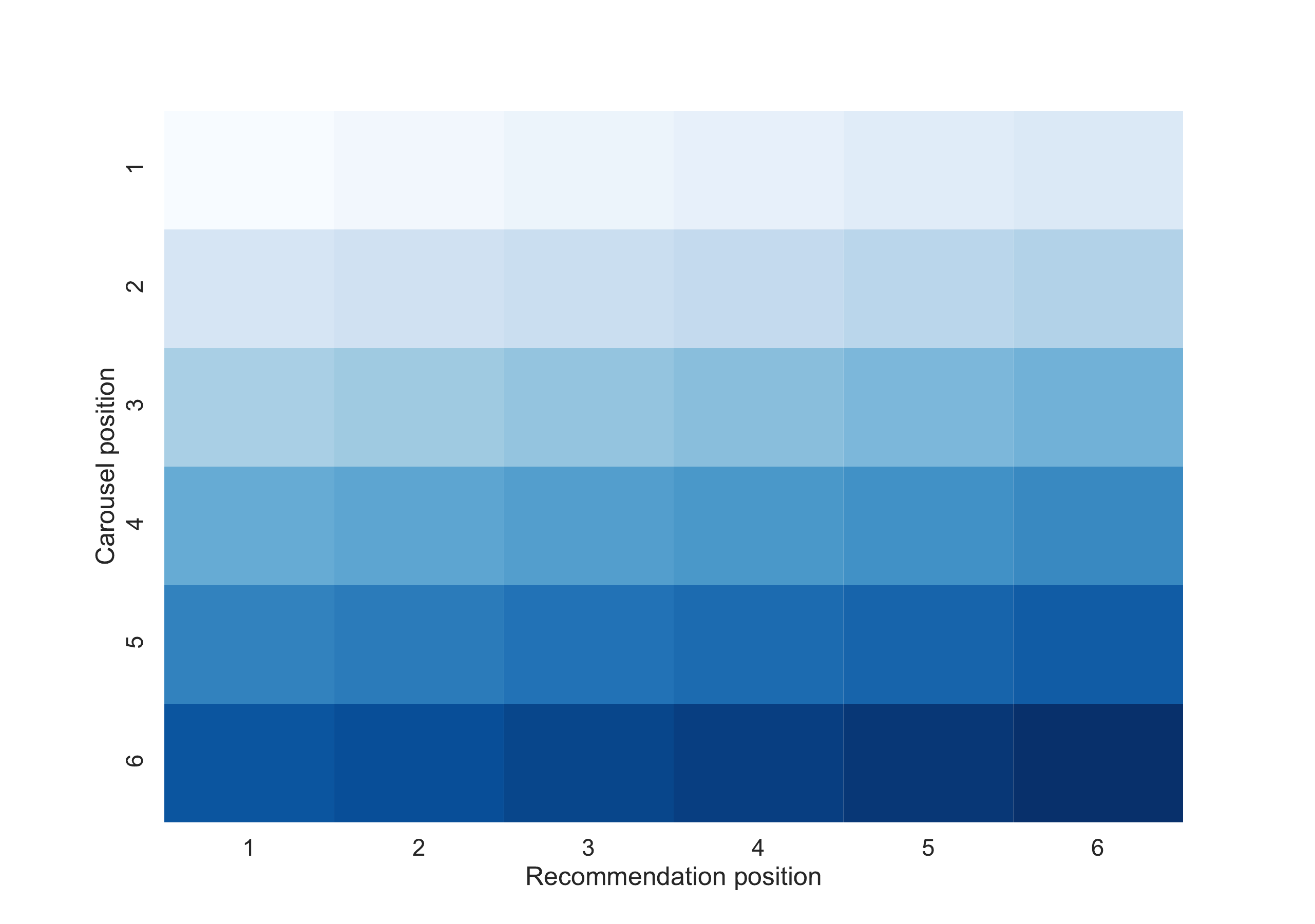}
    \caption{Single list.}
    \label{fig:heatmap_adapted_penalty}
    \end{subfigure}
    \begin{subfigure}{.48\textwidth}
    \includegraphics[width=0.9\columnwidth]{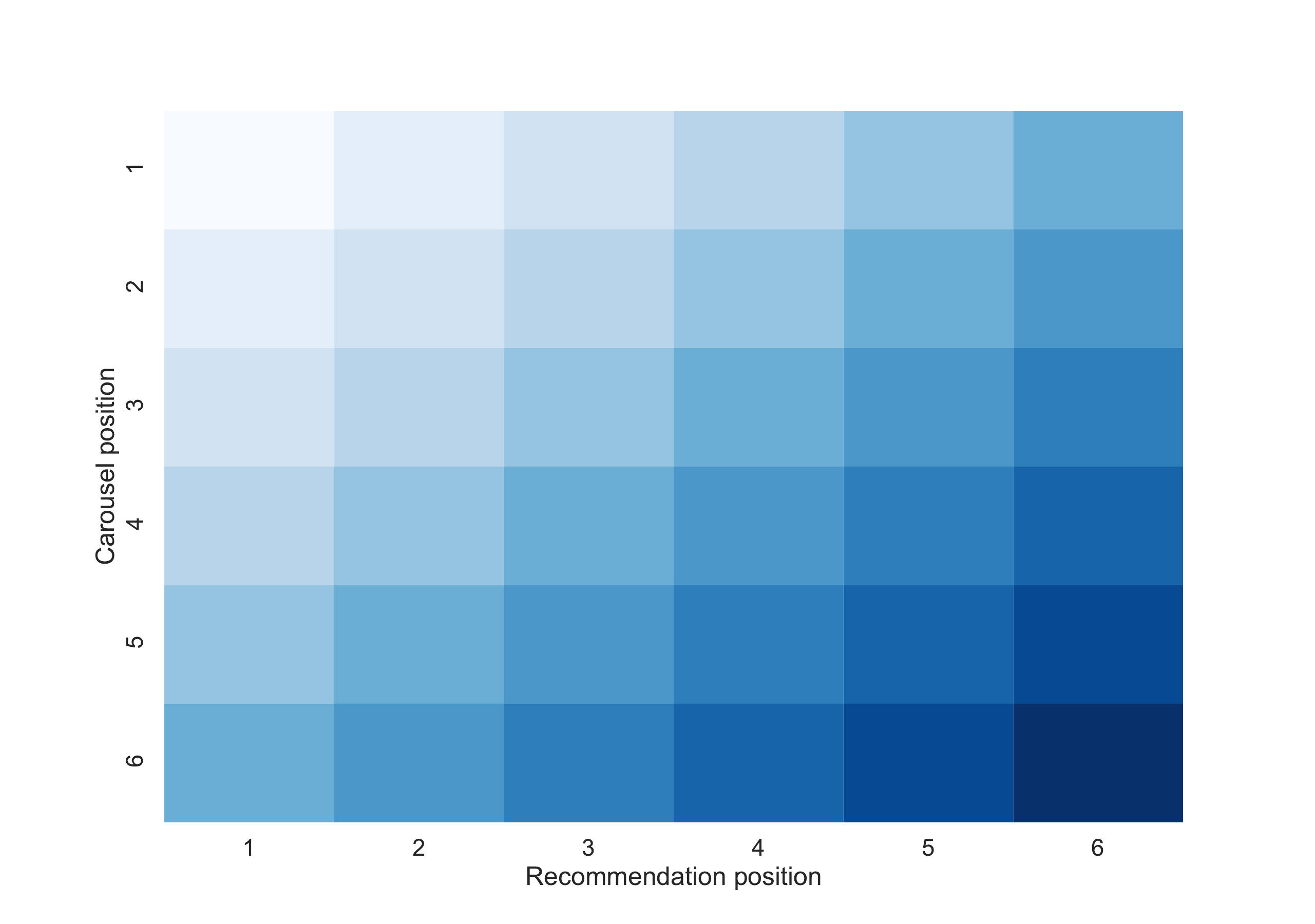}
    \caption{\emph{Golden triangle} behavior.}
    \label{fig:heatmap_triangle_penalty}
    \end{subfigure}
    \caption{A visual comparison of the penalty for each recommended item under different assumptions. Figure \ref{fig:heatmap_adapted_penalty} refers to carousels concatenated in a single list. Figure \ref{fig:heatmap_triangle_penalty} refers to the two-dimensional penalty which accounts for the \emph{golden triangle} behavior.}
    \Description{In this figure, we can see two heatmaps side-by-side. Each heatmap represents a two-dimensional carousel interface, and each element of the grid has a color based on how much that position is penalized in the metric computation. On the left, there is the heatmap corresponding to a metric that interprets the two-dimensional interface as a single list. Therefore, we can see that the penalization follows sequentially the elements as if all the carousels were all concatenated in a single list. On the right, there is the heatmap which considers the golden triangle effect, and so we can see that the penalization is more evident in elements far away from the top-left corner.}
    \label{fig:penalties}
\end{figure*}

One of the most used metrics to measure the quality of a ranked one-dimensional list of results is the \textit{Normalized Discounted Cumulative Gain} (NDCG) \cite{DBLP:conf/sigir/JarvelinK00}. 
This metric is defined based on the DCG, computed as follows:
\begin{equation*}
\label{eq:dcg}
    DCG = \sum_{i=1}^{c} \frac{rel(i)}{log_2(i+1)}  
\end{equation*}

Where $c$ is the length of the recommendation list and $rel(i)$ a function that computes the relevance of the item in position $i$, \eg the rating or just a binary 0-1 value.
In this formulation, the numerator is responsible for rewarding items that are relevant for the user (the function $rel(i)$), while the denominator should penalize the items proportionally to how far they are from the beginning of the list. Clearly, it is better if a correct recommendation is positioned at the beginning of the list rather than at the end.
This reasoning is sound in a layout where a list of results is presented because users will explore the list sequentially \cite{DBLP:conf/etra/KammererG10}. In the case of a two-dimensional layout, this metric could be simply adapted, by assuming that the grid of results will be explored as a long list, with the various rows concatenated one after the other. A visualization of the penalty imposed in this case is given in Figure \ref{fig:heatmap_adapted_penalty}. 

This assumption, though, is not realistic, as many research studies on user behavior suggest \cite{DBLP:conf/etra/KammererG10, DBLP:conf/recsys/ZhaoCHK16}. Hence, we propose to adapt the penalization term to follow the "golden triangle" assumption, i.e., the fact that users tend to explore a two-dimensional interface focusing on the top-left corner, with decreasing attention both to the right and to the bottom.
The penalty that we propose can be computing with the logarithm of the sum of the row and column of each recommendation and is visualized in Figure \ref{fig:heatmap_triangle_penalty}. This penalty can also be modified depending on the particular interface. For example, in a mobile application on a smartphone, we can penalize less the vertical dimension. This can be easily achieved by weighting the row and column coordinates with different weight and therefore can be easily adapted to the desired interface layout.
An important aspect is that duplicate recommendations should be taken into account when computing the numerator. We propose to reward the correct recommendations only the first time the user sees them, but not the following ones others since repeating the same recommendation multiple times not enhance user satisfaction.
With this new definition of the reward and the penalty, we call this metric \textit{NDCG2D}.

\section{Results}
\label{sec:experimental_anaysys}
We now compare the proposed carousel evaluation methodology and the two-dimensional NDCG metric with the traditional evaluation done on a single list at a time. 

We report the results for \textit{MovieLens 10M}, a dataset from the video-on-demand domain, which is among the domains that use the carousel user interface.
MovieLens 10M contains almost 70 thousand users and 10 thousand items, for a total of 10 million ratings in the range 1-5. The dataset also contains item features such as tags, provided by users, the genre of the movie and the year of release.
We randomly selected 80\% of interactions for the training set and 10\% for both validation and test set.

\subsection{Algorithms} 
We select different recommendation algorithms from different families of models.


\begin{description}

    \item[TopPopular:] is a frequently used recommendation technique that recommends to all users the same set of the most popular items. 
    
    \item[ItemKNN Hybrid:] is a nearest-neighbor technique that computes the similarity between items using both the user interactions and the item metadata, using cosine with shrinkage \cite{DBLP:conf/ewmf/MobasherJZ03}.
    
 
    
    \item[\pbetabold:] is a graph-based technique that represents the data as a graph. Users and items are represented as nodes and a connection exists between them if the user has interacted with that item. In \pbeta the similarity between items is computed by simulating a random walk on the graph \cite{DBLP:journals/tiis/PaudelCNB17}.
    

    
    \item[\EASER:] is a machine learning model which was recently proposed \cite{DBLP:conf/www/Steck19}. \EASER is equivalent to a shallow autoencoder, since it has a closed-form solution the model can be computed very rapidly. 

    
    \item[FunkSVD:] is a known matrix factorization algorithm\footnote{\url{https://sifter.org/~simon/journal/20061211.html}} created during the Netflix Prize. FunkSVD decomposes the user-item interactions as the product of two lower dimensional matrices.

    \item[NMF:] Non-negative Matrix Factorization \cite{DBLP:journals/ieicet/CichockiP09} is another matrix factorization algorithm that uses a known matrix-decomposition technique applied on the user-item interactions.

    

\end{description}

In order to ensure a fair comparison between the recommendation quality of all the algorithms, we thoroughly optimized all of them by following the guidelines of \citet{10.1145/3434185}. In particular, we optimized the hyperparameters of all models using a Bayesian search \cite{DBLP:conf/recsys/FelicioniDCBHBD20} exploring 50 cases, the first 15 of them are selected at random while the remaining 35 are selected according to the Gaussian process.\footnote{We relied on the implementation provided in the scikit-optimize python library \url{https://scikit-optimize.github.io/stable/modules/generated/skopt.gp_minimize.html}} We optimized the ranking quality with a recommendation list of 10 elements.

\begin{table*}[]
    \centering
\begin{minipage}{0.6\textwidth}

\resizebox{\columnwidth}{!}{%
    \begin{tabular}{l|c|c|cc|c|}
\toprule
{} & Individual & Carousel & Individual & Carousel &        {} \\
{} &           NDCG &       NDCG 2D &      NDCG rank &  NDCG 2D rank & $ \Delta $ rank \\
\midrule
TopPopular        &         0.0983 &        -- &              -- &             -- &               -- \\
\midrule
\midrule
ItemKNN hybrid &         0.2174 &        0.2148 &              3 &             4 &              -1 \\
\midrule
\pbeta       &         0.2160 &        0.2035 &              4 &             5 &              -1 \\
\midrule
\EASER        &         0.2566 &        0.2293 &              1 &             2 &              -1 \\
\midrule
MF FunkSVD    &         0.2307 &        0.2373 &              2 &             1 &              +1 \\
NMF           &         0.1974 &        0.2281 &              5 &             3 &              +2 \\
\bottomrule
\end{tabular}
}
\end{minipage}
\begin{minipage}{0.39\textwidth}

    \includegraphics[width=\columnwidth]{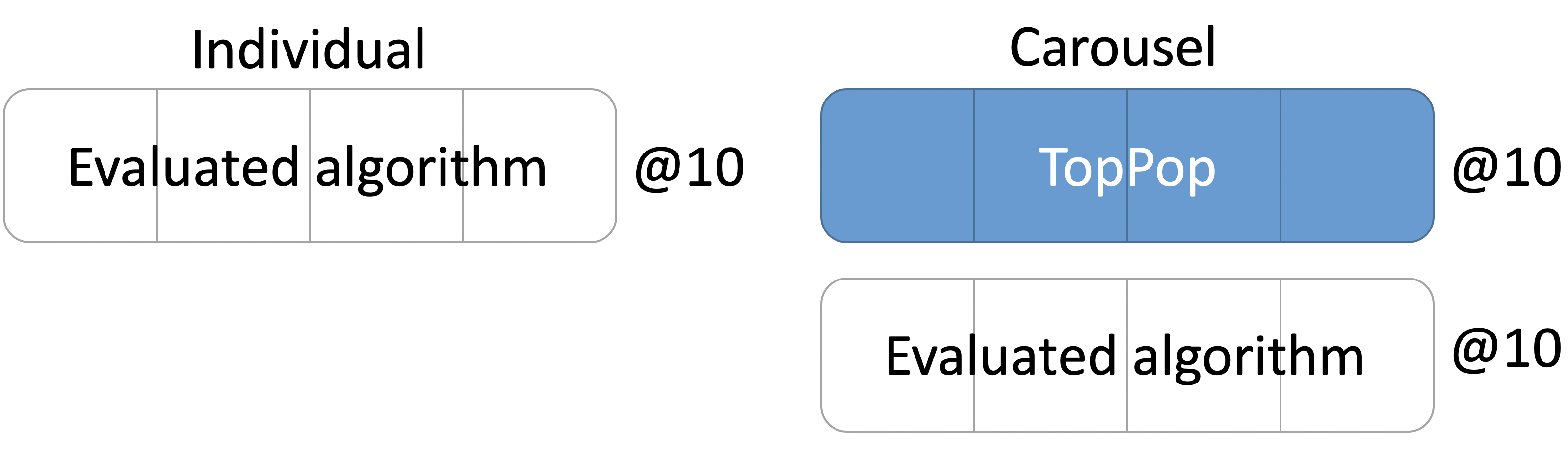}

\end{minipage}
    
    \caption{
    Comparison of the MovieLens10M accuracy metrics with individual and carousel evaluation (with TopPop fixed as the first carousel) at recommendation list length of 10. Note that in the carousel evaluation there will be two recommendation lists.
    }
    \Description{In this table, we can see how the two different evaluation metrics may affect the relative quality of the algorithms considered. These differences are discussed in the text.}
    \label{tab:ml10macc_short}
\end{table*}

\subsection{Discussion}

Each algorithm is evaluated in two settings. In the first setting each model is evaluated independently with the traditional approach. In the second, it is evaluated in a carousel setting where the first carousel is a TopPopular and the evaluated model is the second carousel in the layout (see Table \ref{tab:ml10macc_short}).
In both cases each model generates a recommendation list of 10 elements, but, in the carousel interface two lists are generated, one for each model.


For the two-dimensional NDCG we considered a case where the user explores both vertically and horizontally evenly. Depending on the device and user interface design, in other cases it may be more realistic to attribute a higher importance to the vertical or horizontal exploration.

We report the results in Table \ref{tab:ml10macc_short}, the ranking quality of the models is evaluated with NDCG and NDCG2D.
We can see how there are several differences between the relative performance of the recommender models.
For example, following the individual NDCG, the best model is \EASER, while according to the carousel NDCG2D, FunkSVD is the best. Both matrix factorization techniques tend to gain positions in the NDCG2D evaluation, while all other models lose positions. This result indicate that the algorithm  that exhibited the best recommendation quality taken independently, \EASER, contains recommendations which are already present in a TopPopular and therefore would generate carousels with duplicated recommendations. Matrix factorization models, on the other hand, are less effective on their own but become more effective in a carousel layout because they better complement the recommendations provided by the TopPopular recommender.

These results indicate the importance of taking into account account the user behavior when deciding which is the most appropriate model to be added in a carousel layout in order to ensure the user is provided with the most effective recommendations.

\section{Conclusions and Future Work}
\label{sec:conclusions}
In this paper we proposed a novel approach for the evaluation of the recommendation quality when a carousel user interface is used in which the recommendation quality of a carousel depends on the entire page.

We also extended the NDCG metric to account for the user navigation behavior in a two-dimensional page.

The experimental results indicate that taking into account both the page layout and the user behavior changes the outcome of an offline evaluation leading to prefer a different carousel. This indicates the importance of taking both the page layout and the user behavior into account during the offline evaluation phase.
Important future work is an online study to assess the extent to which this evaluation protocol is able to more closely mirror the user satisfaction, using surveys or A/B testing. In the end, this evaluation protocol could constitute a viable option for the offline evaluation of algorithmic strategies to select the carousel layout of a page in order to provide the user with more effective carousels and improve user satisfaction.

\clearpage

\clearpage
\bibliographystyle{ACM-Reference-Format}
\bibliography{main-references}

\end{document}